# Pt Supported on Plasma-Chemical Titanium Nitride for Efficient Room-Temperature CO Oxidation

E.N. Kabachkov [1,2], E.N. Kurkin [1,2], N.N. Vershinin [1], I.L. Balikhin [1,2], V.I. Berestenko [1], A. Michtchenko [3]\*, Y.M. Shulga [1,4]

[1]*Institute of Problems of Chemical Physics, Russian Academy of Sciences, Chernogolovka 142432, Moscow Region, Russian Federation*

[2]*Chernogolovka Scientific Center, Russian Academy of Sciences, Chernogolovka 142432, Moscow Region, Russian Federation*

[3]*Instituto Politécnico Nacional, SEPI-ESIME-Zacatenco, Av. IPN S/N, Ed.5, 3-r piso, Ciudad de Mexico, C.P. 07738, Mexico*

[4]*National University of Science and Technology MISIS, Leninsky pr. 4, Moscow 119049, Russian Federation*

A B S T R A C T

Catalysts of carbon monoxide oxidation were synthesized by deposition of platinum on titanium nitride (TiN). Two substrates with an average particle size of 18 and 36 nm were obtained by hydrogen reduction of titanium tetrachloride in a stream of microwave plasma of nitrogen. The surface of the catalysts was studied by X-ray photoelectron spectroscopy (XPS). The CO oxidation rate on the 9–15 wt.% Pt loaded TiN catalysts was found to be 120 times higher than that on platinum black with a specific surface of 30 $m^2$/g. Such catalysts are promising for use in catalytic air purification systems.

\*Corresponding author, E-mail address: almitchen@gmail.com (A. Michtchenko)

Titanium nitride (TiN) is widely used due to its hardness, high electrical conductivity, corrosion resistance and high melting point [1–3], as well as its decorative properties, since its reflection spectrum is very similar to the reflection spectrum of gold [4–7]. Recently, TiN has been used as a catalyst (electrocatalyst) for oxygen reduction reactions [8–13], and as a substrate for M/TiN catalysts, where M is a metal [14–21].

Catalytic oxidation of CO has received considerable attention in the related scientific literature (see, for example, publications [22–30] and references to them), due to its wide applications in exhaust gas after-treatment, CO oxidation for proton exchange membrane fuel cells and air purification systems. In this paper we focus on low-temperature oxidation of CO in catalytic air purification systems. We hasten to note that the term "low-temperature oxidation" is rather a tribute to the tradition originating from the work of Haruta et al. [31], who reported that Au can be a highly active catalyst for the oxidation of CO at temperatures below 0 °C. In principle, it is desirable to have catalysts in the air purification systems of residential premises, which work effectively at room temperature (15-25 °C).

In this report, we present data on the synthesis and study of the properties of Pt/TiN catalysts for efficient room-temperature CO oxidation. Nano-sized TiN powder obtained by hydrogen reduction of titanium tetrachloride in a stream of nitrogen plasma was used as a substrate for the preparation of catalysts [32–33]. The study of the properties of catalysts in the oxidation of CO, which is contained in air at low concentrations (less than 100 mg/m$^3$) at 295 K, showed that the CO oxidation rate on the 9–15 wt. % Pt loaded TiN catalysts is 120 times higher than that on platinum black with a specific surface of 30 m$^2$/g.

## EXPERIMENTAL

*Titanium Nitride Synthesis*

Titanium nitride powders were obtained by hydrogen reduction of titanium tetrachloride in a stream of microwave plasma of nitrogen at atmospheric pressure. A mixture of titanium tetrachloride vapors with hydrogen in the required ratio was introduced into a plasma nitrogen stream with a mass-average temperature of about 3000 K, obtained in a p lasmatron using a microwave generator with a frequency of 2450 MHz and maximum useful power of 5 kW.

The particle size of the obtained powder was controlled by changing the flow rate of TiCl$_4$, which was 0.1 g/min in obtaining TiN powder with an average particle size of 18 nm and 0.25 g/min in obtaining powder with an average particle size of 36 nm. Consumption of the plasma-forming nitrogen was 4 m$^3$/h and consumption of the hydrogen was 0.5 m$^3$/h in both cases. Chemical interaction of reagents and condensation of titanium nitride nanoparticles occurred in a tubular reactor with a diameter of 50 mm and a length of 250 mm, the inner walls of which were lined with quartz. The titanium nitride particles which formed in the reactor after cooling the stream were separated from the gas phase by filtration using a bag filter.

The average particle size of titanium nitride was determined by measuring the specific surface of the powders by low-temperature adsorption of molecular nitrogen (BET method). The particle size $l$ was calculated by the formula $S_{ss} = 6/(l\rho)$, where $S_{ss}$ is the specific surface of the powder and $\rho$ is the specific density of titanium nitride.

*Catalyst preparation*

To obtain a catalyst, an aqueous solution of $H_2PtCl_6\ 6H_2O$ ($10^{-2}$ mol/l) was mixed with an aqueous solution of LiCOOH (0.04–0.1 mol/l) at 20 °C. Then, TiN was ultrasonically dispersed in water at 60 °C, followed by addition of the required amount of Pt in the form of an $H_2PtCl_6$/LiCOOH mixture. A similar method has been described in detail earlier [20].

After the induction period (8–15 minutes), platinum clusters precipitated on the surface of titanium nitride. The solution was kept at room temperature for 24 hours, after which the catalyst was washed from the reaction products with distilled water (5–6 times).

Washed catalyst was dried at a temperature of 80 °C for 24 hours. Partial reduction of platinum clusters was then carried out in a $CO-N_2$ atmosphere (volume fraction of CO was 10%) at a temperature of 90 °C for 4 hours. Two catalysts K18 and K36 with a platinum content of 12 wt. % were selected as the main objects of study. TiN of particle size of 18 ± 2 nm and 36 ± 2 nm was used as a substrate in the K18 and K36 catalysts, respectively.

*Sample characterization*

Surface areas of the TiN samples were obtained from $N_2$ sorption isotherms measured at 77 K on QUADRASORB *SI* Analyzer (Quantachrome Instruments). X-ray patterns were recorded using a DRON ADP-2-02 diffractometer with Cu Kα anode (λ = 0.154056 nm). A JEOL JEM 2100 electronic transmission microscope was used to study the structure and composition of Pt/TiN catalysts. An Acculab ALC-80d4 analytical scale was used to weigh the reagents and samples.

XPS spectra were obtained using a Specs PHOIBOS 150 MCD electron spectrometer with an Mg cathode (*hv* = 1253.6 eV). The vacuum in the spectrometer chamber did not exceed $4 \times 10^{-8}$ Pa. The spectra were recorded in the constant transmission energy mode (40 eV for survey spectra and 10 eV for individual lines). The survey spectrum was recorded in 1.00 eV increments, while the spectra of individual lines were recorded in 0.03 eV increments. Background subtraction was carried out according to the Shirley method [34], and spectra decomposition was performed according to the set of mixed Gaussian/Lorentz peaks in the framework of the Casa XPS 2.3.19 software. For quantitative estimates, we used the table values of specific densities (4.24 g/cm$^3$ for $TiO_2$ and 5.44 g/cm$^3$ for TiN), as well as the following values of photoelectron escape depths [35]: $\lambda_1 = \lambda_{Ti2p}^{TiO2} = 3.08$ nm, $\lambda_2 = \lambda_{Ti2p}^{TiN} = 1.73$ nm.

*Method for the study of catalytic properties*

The kinetics of CO oxidation in air on the catalyst was studied according to the method described in detail earlier [36]. In brief, the test chamber was purged for 600 s with a gas mixture of carbon monoxide (150 mg/m$^3$) and air at a speed of 50 cm$^3$/s. The inlet and outlet valves of the test chamber were then closed and the air/gas mixture pump located in the test chamber was turned on, ensuring circulation of the gas mixture through the catalyst at a speed of 30 cm$^3$/s. After the concentration of CO was reduced due to the catalytic reaction to the level of 100 mg/m$^3$, a digital stopwatch was turned on and the readings of the sensors were recorded. The test chamber with a volume of 300 cm$^3$ was equipped with NAP-505

CO sensor (Nemoto), MSH optical sensor $CO_2$ – P/$CO_2$/NC/5/V/P (Dynament), humidity sensor and temperature sensor SHT75 (Sensirion).

RESULTS AND DISCUSSION

*Titanium Nitride*

There are many publications in the literature devoted to TiN, considering that it is of great practical interest, including investigation by XPS method [37–44]. However, the interpretation of experimental data obtained by XPS is somewhat different for different authors. This conclusion relates primarily to quantitative estimates which are connected both with different ways of subtracting the background in the XPS spectra and with the complexity of the object itself.

The fact is that titanium nitride exists as a homogeneous phase over a relatively wide range of compositions and has a tendency to oxidation. The composition and structure of the oxidized layer on the surface of titanium nitride depends on both the preparation method and storage conditions, and on the particle size.

Fig. 1 shows a survey spectrum of one of the samples of plasma-chemical titanium nitride (36 nm). We note immediately that the spectra of other samples are not fundamentally different from those given. Table 1 lists the elemental content (in atomic percent) in the layer analyzed by XPS (2-4 nm). We see that the nitride particles are covered with a thick layer of contamination, the origin of which is associated with the high activity of titanium nitride nanoparticles and the conditions of their sufficiently long storage in air. As for the presence of silicon and sulfur in the sample, these are associated with the technology used for producing titanium nitride.

Table 1. XPS composition of the samples under study

| Sample | Composition (at. %) | | | | | | |
|---|---|---|---|---|---|---|---|
| | C | N | O | Pt | Si | Ti | S |
| TiN (36 nm) | 58.8 | 6.3 | 23.3 | -- | 2.5 | 7.1 | 1.8 |
| К36 | 75.9 | 1.8 | 15.4 | 0.4 | 1.2 | 4.9 | >0.1 |
| К18 | 54.3 | 5.6 | 26.5 | 0.9 | 0.2 | 12.3 | >0.1 |

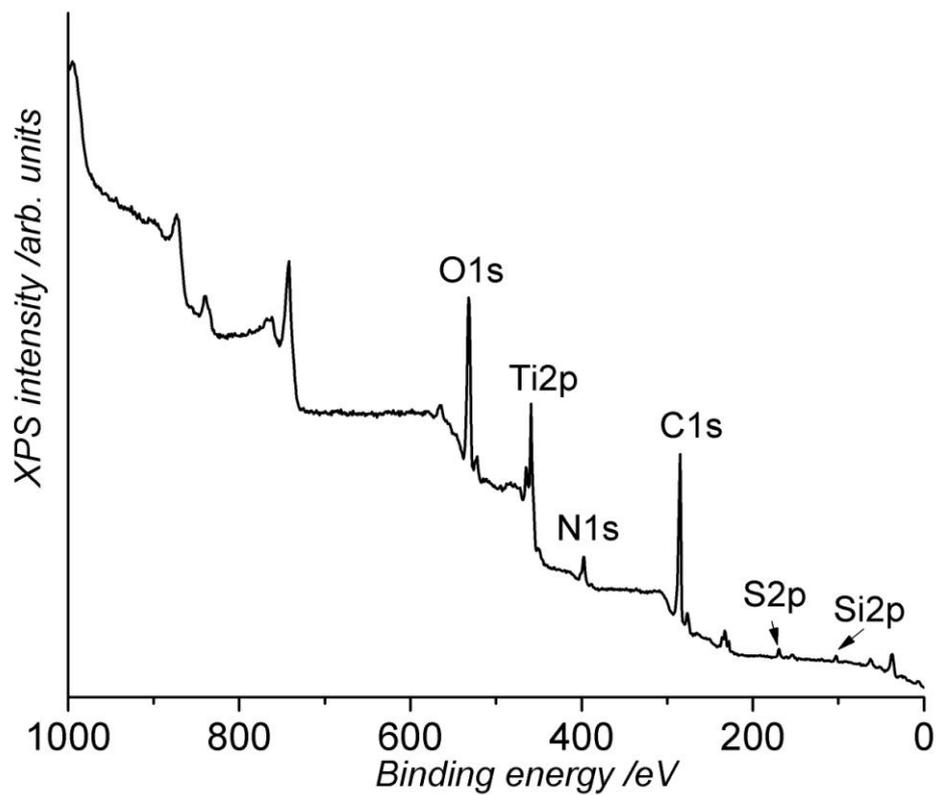

Fig. 1. XPS survey spectrum of the TiN.

Let us analyze the shape of the spectra of Ti2p and N1s. The Ti2p spectrum of an individual titanium compound is known to be a spin-orbit doublet, which is described by two peaks (Ti2p$_{1/2}$ and Ti2p$_{3/2}$) with an intensity ratio of 1: 2 and a distance between the peaks of 5.7 eV [45].

The experimental Ti2p spectrum of our titanium nitride is well described by 6 peaks or 3 doublets (Fig. 2) corresponding to titanium in nitride (1), oxynitride (2) and oxide (3). The positions and relative intensities of the Ti2p$_{3/2}$ peaks are shown in Table 2.

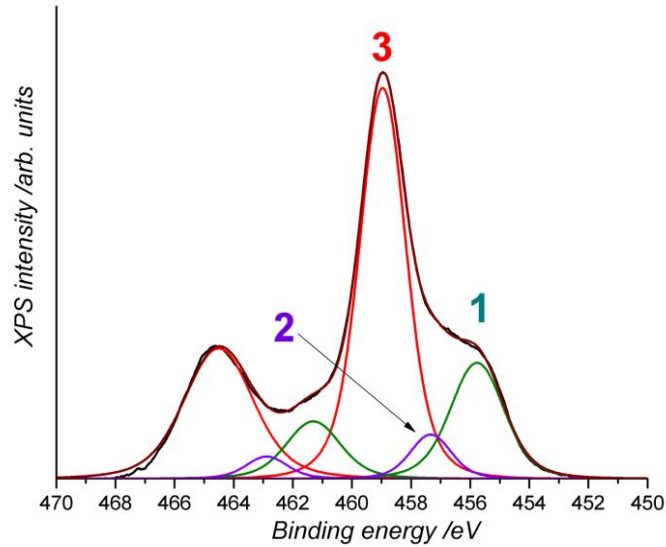

Fig. 2. XPS Ti2p spectrum of the titanium nitride.

Table 2. Peak positions, full width at half-maximum (FWHM) and intensities of the Ti2p$_{3/2}$ peaks obtained by decomposing the Ti2p spectra of the samples under study. For designation of the parameter "$d$", see the text.

| Sample | Peak | $E_b$, eV | FWHM, eV | I, % | $d$, nm |
|---|---|---|---|---|---|
| TiN | 1 | 455.7 | 1.8 | 15.1 | |
| | 2 | 457.3 | 1.7 | 6.9 | 3.7 (1.33) |
| | 3 | 459.0 | 1.8 | 44.7 | |
| K18 | 1 | 456.1 | 2.0 | 30.3 | |
| | 2 | 457.6 | 1.6 | 11.2 | 1.6 (0.71) |
| | 3 | 459.1 | 1.7 | 25.1 | |
| K36 | 1 | 455.8 | 1.9 | 29.4 | |
| | 2 | 457.4 | 1.7 | 13.1 | 1.5 (0.67) |
| | 3 | 458.9 | 2.1 | 24.1 | |

The thickness of the oxide film $d$ can be calculated by a simple formula (see, for example, [34]):

$$\frac{I_3}{I_1} = \frac{I_{Ti2p}^{TiO_2}}{I_{Ti2p}^{TiN}} = \frac{\sigma_{Ti2p}^{TiO_2}}{\sigma_{Ti2p}^{TiN}} \frac{n_{Ti}^{TiO_2}}{n_{Ti}^{TiN}} \frac{\lambda_{Ti2p}^{TiO_2}}{\lambda_{Ti2p}^{TiN}} [exp(d/\lambda_{Ti2p}^{TiO_2}) - 1], \qquad (1)$$

where $I_1$ and $I_3$ are the intensity of the peaks in Table 2, $\sigma_{Ti2p}$ is ionization cross section of Ti2p level, $\lambda_{Ti2p}^{TiO2}$ is the escape depth of Ti2p photoelectrons from the oxide layer on the surface of the nitride.

Formula (1) for nanoparticles gives an overestimated value of $d$ (Table 2), since it is derived for a flat infinite sample coated with an oxide film, and does not take into account the contributions to the intensity $I_3$ of the oxide film from the side surfaces of the TiN nanoparticles.

Calculation of the intensities $I_1$ and $I_3$, carried out for a cubic nanoparticle of titanium nitride coated on all sides with an oxide layer ($l$ – is the edge of the TiN cube, $d$ – is the thickness of the oxide layer on it), assuming that the emission of photoelectrons is recorded in the direction perpendicular to one of the faces of the cube, gives the following expressions:

$$I_1 = (l - 2d)^2 \, \sigma_{Ti2p}^{TiN} \, n_{Ti}^{TiN} \, \lambda_{Ti2p}^{TiN} \, exp(-d/\lambda_{Ti2p}^{TiN}) \, exp(-d/\lambda_{Ti2p}^{TiO_2}), \qquad (2)$$

$$I_3 = \sigma_{Ti2p}^{TiO_2} \, n_{Ti}^{TiO_2} \, \lambda_{Ti2p}^{TiO_2} [l^2 - (l - 2d)^2 \, exp(-d/\lambda_{Ti2p}^{TiO_2})], \qquad (3)$$

Estimation of the values of $d$ by Eqs (2) and (3) gives values of 0.71 nm and 0.67 nm for K18 and K36, respectively. These estimates are shown in Table 2 in parentheses. Note that the calculation by formulas (2) and (3), in accordance with the simplification carried out, gives underestimated values of the oxide layer thickness.

In the spectrum of N1s (Fig. 3), in addition to the main peak related to nitrogen in the mononitride lattice ($E_b$ = 397.1 eV) (see, for example, [37, 45]), we can also distinguish 2 peaks with $E_b$ = 399.2 and 401.6 eV (Table 3). According to data from the literature [46–47], the peak with $E_b$ = 399.2 eV can be associated with nitrogen atoms in the lattice of the oxynitride Ti(N,O). The peak with $E_b$ = 401.6 eV is often attributed in the literature to molecular nitrogen formed during the oxidation of nitride [47–49]. However, in the XPS spectra of transition metal (M) complexes with molecular nitrogen (M-$N_2$), the N1s line is split into 2 peaks [50–54]. If we assume that $N_2$ is coordinated in the same way as in a binuclear complex (M-N-N-M), then the value of $E_b$ (N1s) should be the same as for the *exo* atom of the mononuclear complex, i.e., below the specified value by at least 1 eV. In our opinion, the origin of the peak with $E_b$ = 401.6 eV is still questionable. Moreover, many authors have not noted this peak in the XPS spectra of the samples of titanium nitride studied by them. In our opinion, adsorbed NO molecules, nitrogen atoms in the $TiO_2$ lattice and N1s photoelectrons from nitride nitrogen, which have lost some of their energy due on the excitation caused by electron transitions from the conduction band to the free band, can contribute to photoemission in the region near 401.6 eV.

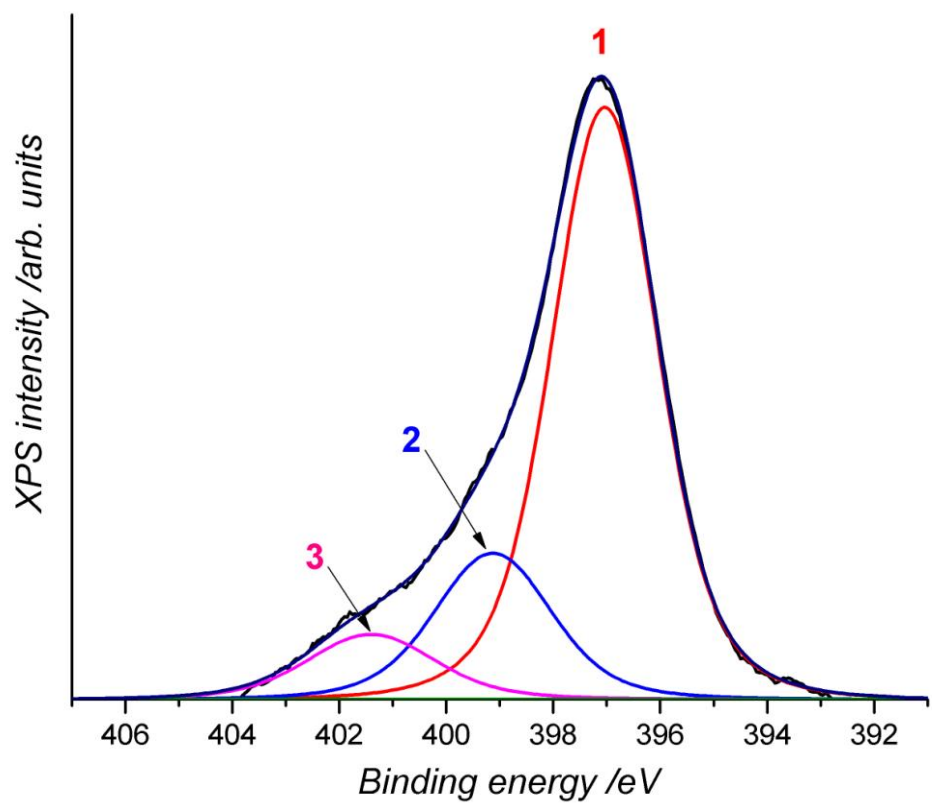

Fig.3. N1s XPS spectrum of the titanium nitride.

Table 3. Positions, half-widths and intensities of peaks obtained by decomposing the N 1s spectra of the initial titanium nitride and catalysts.

| Sample | Peak | $E_b$, eV | FWHM, eV | I, % |
|---|---|---|---|---|
| TiN | 1 | 397.1 | 2.3 | 71.3 |
|  | 2 | 399.2 | 2.5 | 19.3 |
|  | 3 | 401.6 | 2.8 | 9.2 |
| K18 | 1 | 397.1 | 2.0 | 18.1 |
|  | 2 | 399.4 | 2.6 | 80.1 |
|  | 3 | 401.6 | 1.9 | 1.7 |
| K36 | 1 | 397.1 | 2.7 | 30.8 |
|  | 2 | 399.4 | 2.3 | 61.4 |
|  | 3 | 401.6 | 2.7 | 7.7 |

Thus, the particles of initial titanium nitride are covered with a rather thick film of titanium oxide, which contains nitrogen atoms. Between the nitride and the oxide on its surface is a thin layer of oxynitride.

The stability of bulk samples of titanium nitride in an oxidizing environment is well known. It is clear that the density of titanium atoms in $TiO_2$ is noticeably lower than that in TiN (data on the specific density of $TiO_2$ and TiN are given above). Therefore, $TiO_2$ on the surface of TiN cannot serve as a protective film preventing diffusion of oxygen to titanium nitride. However, many authors claim that the oxide film on the surface of titanium nitride consists of pure $TiO_2$ [55–56].

The data obtained by us in the present work and earlier [57–59] indicate the presence of oxynitride as a transition layer between nitride and oxide. It is this layer that constitutes the barrier preventing the oxidation of nitride. This data also coincides with the conclusions of Ref. [60].

*Catalysts*

The contents of elements (in atomic percent) in the near-surface layer of catalysts are presented in Table 1. As in the initial TiN, a rather thick hydrocarbon film can be seen on the catalyst surface (high carbon content). The reasons for the appearance of hydrocarbon contamination of the surface are listed above. It may be noted that in the process of preparing catalysts, the $[N/Ti]_{at}$ ratio, calculated from the integrated intensities of the N1s and Ti2p lines, decreases by a factor of 2.

The results of decomposition of the XPS spectra of Ti2p catalysts are presented in Table 2. It can be seen that the ratio of $I_3/I_1$ in catalysts is lower than in the initial titanium nitride. Consequently, the thickness of the oxide film on the surface of TiN in the catalyst is less than that in the initial TiN. How could this happen?

It can be assumed that the reduction of surface titanium oxide also occurred during the reduction of platinum. However, a second question arises—in what form is the reduced part of the titanium oxide film present in the catalyst? Obviously, if the reduction to metal occurred, then contact with air will again lead to oxidation of the metal and formally nothing should change. It seems to us that ultrasonic mixing and subsequent washing with water cause the changes in the I3/I1 ratio during the deposition of platinum. In these operations, it is likely that the top friable layer of the oxide film gets mechanically destroyed and the separated fine oxide particles are removed from the sample during washing. Analyzing the data of Table 2, it should also be noted that the intensity of peak 2 ($I_2$) in the catalysts is higher than that in the initial sample.

On the N1s spectrum (Fig. 4) of the catalyst, the main peak is the peak with $E_b$ = 399.2 eV. A synchronous increase in the intensities from the oxynitride layer in the N1s and Ti2p spectra means that the influence of this surface layer on the electronic properties of the surface has increased significantly.

The direct contact of platinum with oxynitride cannot be excluded either, since the oxide layer does not have to be continuous. Consequently, the properties of the contact between the catalytically active metal (Pt) and the substrate in the catalyst under study may be differ significantly from those the Pt/TiO$_2$ contact.

The spectrum of Pt4f is well described by two doublets Pt4f$_{7/2}$ and Pt4f$_{5/2}$ (Fig. 5), one of which, with $E_b$ (Pt4f$_{7/2}$) = 71.4 eV, corresponds in its position to metallic platinum, and the second (with $E_b$ (Pt4f$_{7/2}$) = 74.3 eV) to Pt$^{4+}$ oxide. It should be noted here that the treatment of the catalyst with carbon monoxide does not lead to the complete reduction of platinum.

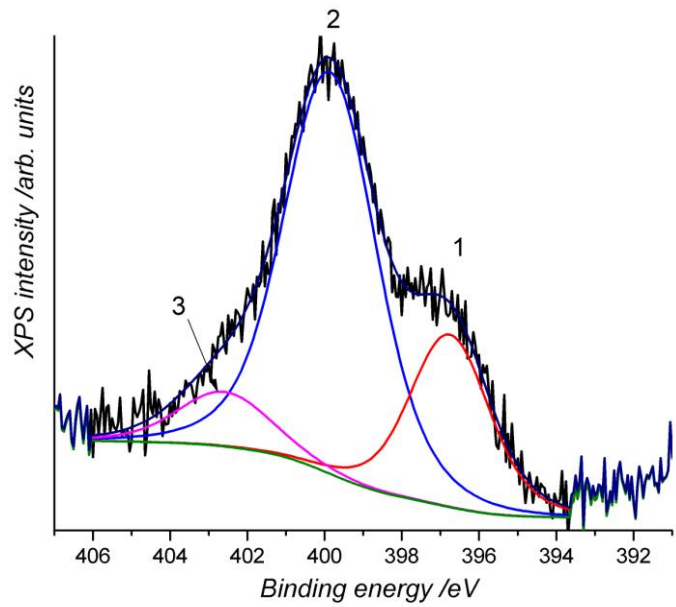

Fig.4. XPS N1s spectrum of catalyst K32.

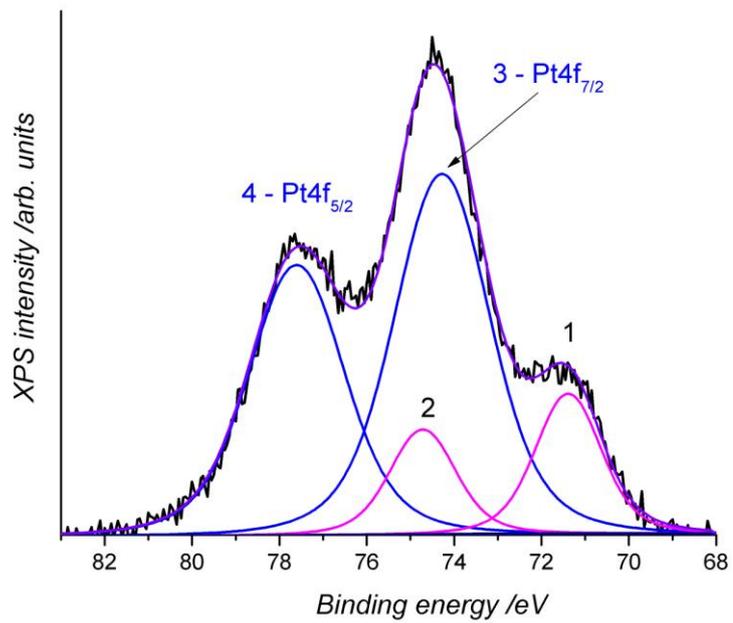

Fig. 5. XPS Pt4f spectrum of K32 catalyst after treatment with carbon monoxide.

We also note here that the intensity of the Pt4f line is noticeably lower than would be expected from the assumption of a homogeneous distribution of 12 mass. % platinum (Table 1). If we recalculate the surface composition of catalysts from atomic percent to mass, as given in this Table, then of course the proportion of platinum will increase; however, for both studied catalysts it will remain less than 12 mass %. The reason for this may be that the distribution of platinum clusters is not homogeneous. Such a design naturally suppresses the output of Pt4f photoelectrons. For the system of platinum on $TiO_2$, the phenomenon of strong metal-support interaction (SMSI) is often observed when the metal is covered with an oxide film [61–62].

*X-ray patterns*

The titanium nitride obtained by us was found to have an NaCl type lattice. After deposition of platinum, the lattice parameter of titanium nitride does not change. Fig. 6 shows the diffraction patterns of catalysts. First, we note that the values of the full width at half maximum (FWHM) of the reflex Pt (111) in the X-ray patterns of the samples under study are more than those for TiN (111). This means that the sizes of platinum clusters are smaller than the particle sizes of the substrate. Using Scherrer's equation, the width of the Pt(111) line gives the coherence length $L_c = 8$ nm for K18. For K36, $L_c = 12$ nm.

*TEM*

Fig. 7 shows a surface image of catalyst K36. Platinum clusters with a size of 4–5 nm can be seen on the surface of titanium nitride particles. However, the distribution of platinum clusters cannot be called uniform. In the figure, you can see titanium nitride particles on which there are no Pt clusters. The presence of a large number of platinum clusters on some particles of titanium nitride and their absence on other particles of titanium nitride leads to a decrease in the surface concentration of platinum as determined by the XPS method.

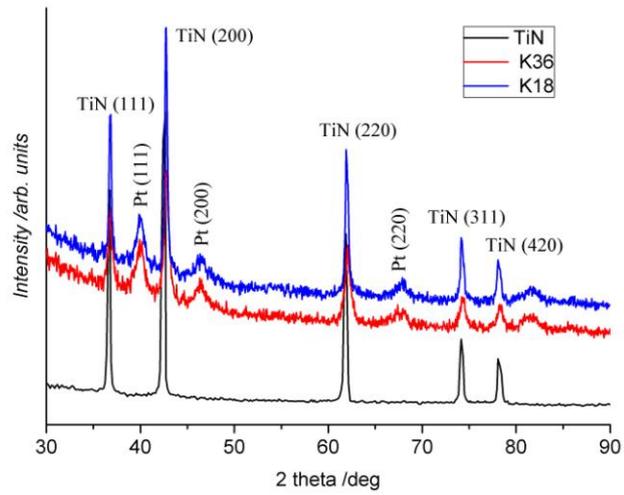

Fig. 6. X-ray patterns of TiN and K18 and K36 catalysts

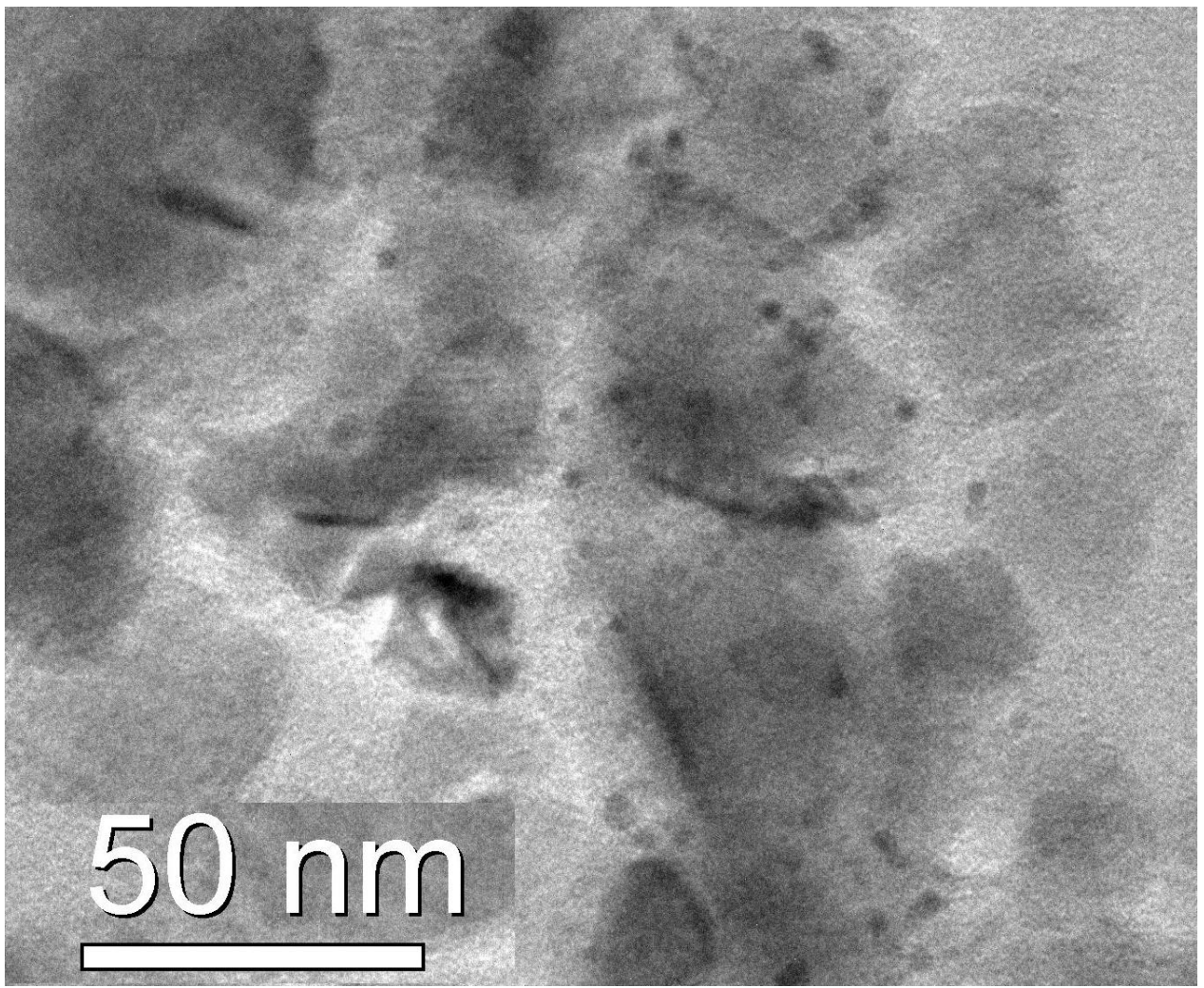

Fig. 7. TEM image of Pt clusters on the surface of TiN (36 nm)

*Catalytic properties of Pt/TiN samples*

An increase in $CO_2$ concentration occurs at the same time as the decrease in the CO concentration in the test chamber, due to the irreversible oxidation reaction of CO with air oxygen. As can be seen from Fig. 8, the time dependence of the CO concentration in the test chamber with a catalyst is described by equation (4):

$$C_{CO}(t) = C_{CO}(0)\, e^{-k\,t}, \qquad (4)$$

where $C_{CO}(t)$ is the measured value of the concentration of CO in the test chamber, $C_{CO}(0)$ is the value of concentration of CO at the initial (zero) time, $k$ is the reaction rate constant, $t$ is the time.

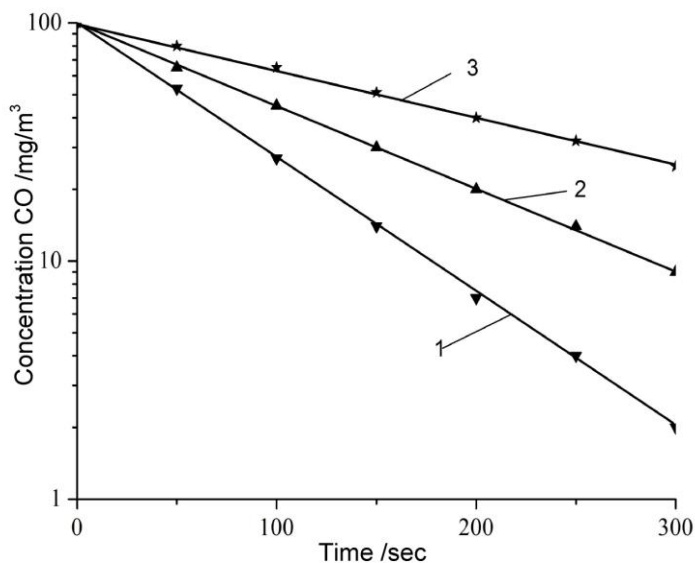

Fig. 8. Kinetics of CO oxidation at T = 295 K, P = 101 kPa and 30% humidity on K18 (1), K36 (2) catalysts and on Pt black (3) (at the same platinum content in the samples).

The experimental data in Fig. 8 reveal that the reaction rate constant on K18 is 120 ± 20 times higher than that of Pt black with a specific surface of 30 $m^2$/g. It should also be noted that the reaction rate constant on K18 is 1.5 ± 0.1 times higher than that on K36.

To study the effect of platinum content on catalytic properties, five samples of a catalyst based on titanium nitride with a particle size of 18 nm were synthesized and tested (Table 4). The investigation of the CO oxidation reaction of these samples was carried out according to the procedure described above.

Table 4. The dependence of the catalytic properties of Pt/TiN on Pt content at T = 295 K, P = 101 kPa, RH = 30%

| No sample | $V_x/V_{12}$ | Sample weight, g | Pt content, mass % |
|---|---|---|---|
| 1 | 0.9 | 0.050 | 6 |
| 2 | 0.95 | 0.033 | 9 |
| 3 | 1.0 | 0.025 | 12 |
| 4 | 0.95 | 0.020 | 15 |
| 5 | 0.7 | 0.012 | 25 |

When studying the catalytic properties, the mass of the catalyst was changed depending on the composition, leaving the same mass of platinum in each sample equal to 3 ± 0.3 mg. The reaction rates of CO oxidation for each sample were determined after processing the experimental data in accordance with equation (3). The ratio of reaction rates $V_x/V_{12}$ was then determined, where $V_{12}$ is the CO oxidation rate for a composition containing 12 mass % of platinum with the maximum rate of oxidation of CO, and $V_x$ the rate of oxidation of CO for a catalyst containing $x$ mass % Pt. The experimental results presented in Table 4 suggest that the oxidation rate of CO varies little when the platinum content is within an interval of 9 to 15 mass %.

To determine the adsorption properties of platinum in the catalysts and in platinum black (90 mg), the absorption of CO from the gas phase was measured. The test chamber was flushed with dry nitrogen at a rate of 50 cm$^3$/s for 100 s, and then purged with a mixture of CO and N$_2$ for 5 s at a rate of 50 cm$^3$/s. Then the inlet and outlet valves were closed and after 10 min the content of the volume fraction of CO was analyzed. For the analysis of CO, 3 cm$^3$ samples were taken from the test chamber, and then the sample was injected into a measuring chamber filled with air. The number of CO molecules adsorbed by the Pt surface per unit mass of platinum was calculated using formula (5):

$$N_{CO} = (C_{ic} - C_{fc})\, V_o\, N_A / (100\, m_{Pt}\, V_M) \qquad (5)$$

where $C_{ic}$ is the initial volume fraction of CO in %, $C_{fc}$ is the final volume fraction of CO in %, $V_o$ is the volume of the test chamber, $N_A$ is the Avogadro number, $m_{Pt}$ is the mass of platinum in the catalyst in grams, $V_M$ is the molar gas volume. The mass of the catalyst was chosen so as to ensure that the value of the final concentration remained in the range from 0.45 to 0.55 of the initial volume fraction of CO equal to 1%.

After processing the experimental results, the ratio of $N_{CO}$(Pt/TiN)/$N_{CO}$(Pt-black) was found to be equal to 3.8 ± 0.4 for K18 and 2.5 ± 0.3 for K36. The measurements were carried out at T = 295 K and P = 101 kPa.

Thus, an increase in the reaction rate of CO oxidation on Pt/TiO$_2$/TiN catalysts as compared to platinum black can be associated with both an increase in the concentration of CO molecules adsorbed and a decrease in the activation energy of the reaction. The reaction rate increases only by 3.8 ± 0.4 times due to increase in the concentration of CO molecules on the surface of platinum. Therefore, the main reason for the increase in the reaction rate is possibly associated with a decrease in the activation energy of the CO oxidation reaction. From the Arrhenius equation, it follows that with a 32 ± 5 times increase in the reaction rate, the calculated value of the decrease in activation energy will be from 8.1 kJ/mol to 8.5 kJ / mol. The reason for the decrease in the activation energy can be, for example, the formation of oxide layers of platinum on the surface of the Pt cluster [63] or the influence of support (as in the case of support doping [64]).

*Conclusions*

New catalysts of carbon monoxide oxidation were synthesized by deposition of platinum on titanium nitride with an average particle size of 18 nm and 36 nm. It was established that:

1) as a result of catalyst synthesis, the oxide film on the surface of titanium nitride is enriched with nitrogen, and its thickness decreases;

2) surface content of Pt is less than volume content;

3) treatment of the catalyst with carbon monoxide does not lead to the complete reduction of platinum.

The catalytic properties of Pt/TiN samples in the oxidation of CO at room temperature and low CO concentrations (less than 100 mg/m$^3$) were studied. The reaction rate constant of CO oxidation on the 9–15 wt.% Pt loaded TiN catalysts was found to be 120 times higher than that on platinum black with a specific surface of 30 m$^2$/g.

The developed Pt/TiN catalyst is promising for use in catalytic and photo-catalytic air purification devices at low CO concentrations, subject to further research.

*Acknowledgements*

This work was performed in accordance with the state tasks, state registration Nos AAAA-A19-119061890019-5 and 0089-2019-0012. The work has been performed using the equipment of the Multi-User Analytical Center of IPCP RAS and Chernogolovka Scientific Center. This study was carried out with the use of resources of Competence Center of National Technology Initiative in IPCP RAS.

# References

[1] J.-E. Sundgren, J. -E. Sundgren, B. -O. Johansson, S. -E. Karlsson, H.T.G. Hentzell, Mechanisms of reactive sputtering of titanium nitride and titanium carbide II: Morphology and structure, Thin Solid Films. 105 (1983) 367–384. doi:10.1016/0040-6090(83)90319-x.

[2] W.D. Sproul, P.J. Rudnik, C.A. Gogol, The effect of target power on the nitrogen partial pressure level and hardness of reactively sputtered titanium nitride coatings, Thin Solid Films. 171 (1989) 171–181. doi:10.1016/0040-6090(89)90042-4.

[3] H.-E. Cheng, Y.-W. Wen, Correlation between process parameters, microstructure and hardness of titanium nitride films by chemical vapor deposition, Surface and Coatings Technology. 179 (2004) 103–109. doi:10.1016/s0257-8972(03)00789-8.

[4] B. Zega, M. Kornmann, J. Amiguet, Hard decorative TiN coatings by ion plating, Thin Solid Films. 45 (1977) 577–582. doi:10.1016/0040-6090(77)90249-8.

[5] R. Buhl, H.K. Pulker, E. Moll, TiN coatings on steel, Thin Solid Films. 80 (1981) 265–270. doi:10.1016/0040-6090(81)90233-9.

[6] Mumtaz, W.H. Class, Color of titanium nitride prepared by reactive dc magnetron sputtering, Journal of Vacuum Science and Technology. 20 (1982) 345–348. doi:10.1116/1.571461.

[7] S. Niyomsoan, W. Grant, D.L. Olson, B. Mishra, Variation of color in titanium and zirconium nitride decorative thin films, Thin Solid Films. 415 (2002) 187–194. doi:10.1016/S0040-6090(02)00530-8.

[8] D.H. Youn, G. Bae, S. Han, J.Y. Kim, J.-W. Jang, H. Park, S.H. Choi, J.S. Lee, A highly efficient transition metal nitride-based electrocatalyst for oxygen reduction reaction: TiN on a CNT–graphene hybrid support, J. Mater. Chem. A Mater. Energy Sustain. 1 (2013) 8007–8015. doi:10.1039/C3TA11135K.

[9] H. Nan, D. Dang, X.L. Tian, Structural engineering of robust titanium nitride as effective platinum support for the oxygen reduction reaction, J. Mater. Chem. A Mater. Energy Sustain. 6 (2018) 6065–6073. doi:10.1039/C8TA00326B.

[10] Z. Pan, Y. Xiao, Z. Fu, G. Zhan, S. Wu, C. Xiao, G. Hu, Z. Wei, Hollow and porous titanium nitride nanotubes as high-performance catalyst supports for oxygen reduction reaction, J. Mater. Chem. A Mater. Energy Sustain. 2 (2014) 13966–13975. doi:10.1039/C4TA02402H.

[11] Y. Dong, Y. Wu, M. Liu, J. Li, Electrocatalysis on Shape-Controlled Titanium Nitride Nanocrystals for the Oxygen Reduction Reaction, ChemSusChem. 6 (2013) 2016–2021. https://doi.org/10.1002/cssc.201300331.

[12] M. Liu, Y. Dong, Y. Wu, H. Feng, J. Li, Titanium Nitride Nanocrystals on Nitrogen-Doped Graphene as an Efficient Electrocatalyst for Oxygen Reduction Reaction, Chem.--Eur. J. 19 (2013) 14781–14786. https://doi.org/10.1002/chem.201302425

[13] H. Shin, H.-I. Kim, D.Y. Chung, J.M. Yoo, S. Weon, W. Choi, Y.-E. Sung, Scaffold-Like Titanium Nitride Nanotubes with a Highly Conductive Porous Architecture as a Nanoparticle Catalyst Support for Oxygen Reduction, ACS Catal. 6 (2016) 3914–3920. doi:10.1021/acscatal.6b00384.

[14] S. Yang, D.Y. Chung, Y.-J. Tak, J. Kim, H. Han, J.-S. Yu, A. Soon, Y.-E. Sung, H. Lee, Electronic structure modification of platinum on titanium nitride resulting in enhanced catalytic activity and durability for oxygen reduction and formic acid oxidation, Appl. Catal. B. 174-175 (2015) 35–42. doi:10.1016/j.apcatb.2015.02.033.

[15] J. Zhang, L. Ma, M. Gan, S. Fu, Y. Zhao, TiN@ nitrogen-doped carbon supported Pt nanoparticles as high-performance anode catalyst for methanol electrooxidation, J. Power Sources. 324 (2016) 199–207. https://doi.org/10.1016/j.jpowsour.2016.05.083.

[16] J.-M. Lee, S.-B. Han, Y.-J. Song, J.-Y. Kim, B. Roh, I. Hwang, W. Choi, K.-W. Park, Methanol electrooxidation of Pt catalyst on titanium nitride nanostructured support, Appl. Catal. A. 375 (2010) 149–155. doi:10.1016/j.apcata.2009.12.037.

[17] M. Yang, Z. Cui, F.J. DiSalvo, Mesoporous titanium nitride supported Pt nanoparticles as high performance catalysts for methanol electrooxidation, Phys. Chem. Chem. Phys. 15 (2013) 1088–1092. doi:10.1039/c2cp44215a.

[18] M.M.O. Thotiyl, S. Sampath, Electrochemical oxidation of ethanol in acid media on titanium nitride supported fuel cell catalysts, Electrochim. Acta. 56 (2011) 3549–3554. doi:10.1016/j.electacta.2010.12.091.

[19] Y. Xiao, Z. Fu, G. Zhan, Z. Pan, C. Xiao, S. Wu, C. Chen, G. Hu, Z. Wei, Increasing Pt methanol oxidation reaction activity and durability with a titanium molybdenum nitride catalyst support, J. Power Sources. 273 (2015) 33–40. doi:10.1016/j.jpowsour.2014.09.057.

[20] M.M. Ottakam Thotiyl, T. Ravikumar, S. Sampath, Platinum particles supported on titanium nitride: an efficient electrode material for the oxidation of methanol in alkaline media, J. Mater. Chem. 20 (2010) 10643–10651. doi:10.1039/C0JM01600D.

[21] M.Á. Centeno, I. Carrizosa, J.A. Odriozola, Deposition--precipitation method to obtain supported gold catalysts: dependence of the acid--base properties of the support exemplified in the system TiO2--TiOxNy--TiN, Appl. Catal. A. 246 (2003) 365–372. https://www.sciencedirect.com/science/article/pii/S0926860X03000589.

[22] U. Heiz, A. Sanchez, S. Abbet, W.-D. Schneider, Catalytic Oxidation of Carbon Monoxide on Monodispersed Platinum Clusters: Each Atom Counts, J. Am. Chem. Soc. 121 (1999) 3214–3217. doi:10.1021/ja983616l.

[23] N. Lopez, J.K. Nørskov, Catalytic CO oxidation by a gold nanoparticle: a density functional study, J. Am. Chem. Soc. 124 (2002) 11262–11263. https://www.ncbi.nlm.nih.gov/pubmed/12236728.

[24] B.K. Min, C.M. Friend, Heterogeneous gold-based catalysis for green chemistry: low-temperature CO oxidation and propene oxidation, Chem. Rev. 107 (2007) 2709–2724. doi:10.1021/cr050954d.

[25] L. Di, W. Xu, Z. Zhan, X. Zhang, Synthesis of alumina supported Pd–Cu alloy nanoparticles for CO oxidation via a fast and facile method, RSC Advances. 5 (2015) 71854–71858. doi:10.1039/c5ra13813b.

[26] T. Akita, Y. Maeda, M. Kohyama, Low-temperature CO oxidation properties and TEM/STEM observation of Au/γ-Fe2O3 catalysts, J. Catal. 324 (2015) 127–132. doi:10.1016/j.jcat.2015.02.006.

[27] J. He, D. Chen, N. Li, Q. Xu, H. Li, J. He, J. Lu, Hollow Mesoporous Co3O4--CeO2 Composite Nanotubes with Open Ends for Efficient Catalytic CO Oxidation, ChemSusChem. 12 (2019) 1084–1090. https://onlinelibrary.wiley.com/doi/abs/10.1002/cssc.201802501.

[28] Y. Wang, D. Yang, S. Li, L. Zhang, G. Zheng, L. Guo, Layered copper manganese oxide for the efficient catalytic CO and VOCs oxidation, Chem. Eng. J. 357 (2019) 258–268. doi:10.1016/j.cej.2018.09.156.

[29] K. Taira, H. Einaga, The Effect of SO2 and H2O on the Interaction Between Pt and TiO 2 (P-25) During Catalytic CO Oxidation, Catal. Letters. 149 (2019) 965–973. https://doi.org/10.1007/s10562-019-02672-3.

[30] D. Bikaljevic, R. Rameshan, N. Köpfle, T. Götsch, E. Mühlegger, R. Schlögl, S. Penner, N. Memmel, B. Klötzer, Structural and kinetic aspects of CO oxidation on ZnOx-modified Cu surfaces, Appl. Catal. A. 572 (2019) 151–157. doi:10.1016/j.apcata.2018.12.032.

[31] M. Haruta, N. Yamada, T. Kobayashi, S. Iijima, Gold catalysts prepared by coprecipitation for low-temperature oxidation of hydrogen and of carbon monoxide, J. Catal. 115 (1989) 301–309. doi:10.1016/0021-9517(89)90034-1.

[32] V.N. Troitsky, S.V. Gurov, V.I. Berestenko, Peculiarities of production of highly dispersed powders of nitrides of iV group metals by reduction of chloride in low temperature plasma, High Energy Chem. 13 (1979) 267-272.

[33] I.L. Balikhin, V.I. Berestenko, I.A. Domashnev, E.N. Kurkin, V.N. Troitskij, Patent: Installation and method for production of nanodispersed powders in microwave plasma, No 2252817, (2005). https://russianpatents.com/patent/225/2252817.html

[34] J.E. Castle, Practical surface analysis by Auger and X-ray photoelectron spectroscopy. D. Briggs and M. P. Seah (Editors). John Wiley and Sons Ltd, Chichester, 1983, 533 pp., £44.50, Surface and Interface Analysis. 6 (1984) 302–302. doi:10.1002/sia.740060611.

[35] M.P. Seah, W.A. Dench, Quantitative electron spectroscopy of surfaces: A standard data base for electron inelastic mean free paths in solids, Surface and Interface Analysis. 1 (1979) 2–11. doi:10.1002/sia.740010103.

[36] N.N. Vershinin, V.A. Bakaev, V.I. Berestenko, O.N. Efimov, E.N. Kurkin, E.N. Kabachkov, Synthesis and properties of a platinum catalyst supported on plasma chemical silicon carbide, High Energy Chem. 51 (2017) 46–50. doi:10.1134/S0018143916060199.

[37] Y.M. Shulga, V.N. Troitskii, M.I. Aivazov, Y.G. Borodko, X-ray photo-electron spectra of the mononitrides of Sc, Ti, V and Cr, Zh. Neorg. Khim. 21 (1976) 2621–2624.

[38] L.I. Johansson, P.M. Stefan, M.L. Shek, A. Nørlund Christensen, Valence-band structure of TiC and TiN, Physical Review B. 22 (1980) 1032–1037. doi:10.1103/physrevb.22.1032.

[39] L. Porte, L. Roux, J. Hanus, Vacancy effects in the x-ray photoelectron spectra ofTiNx, Physical Review B. 28 (1983) 3214–3224. doi:10.1103/physrevb.28.3214.

[40] Bertóti, M. Mohai, J.L. Sullivan, S.O. Saied, Surface characterisation of plasma-nitrided titanium: an XPS study, Applied Surface Science. 84 (1995) 357–371. doi:10.1016/0169-4332(94)00545-1.

[41] D. Gall, R.T. Haasch, N. Finnegan, T. -Y. Lee, C. -S. Shin, E. Sammann, J.E. Greene, I. Petrov, In situ X-ray Photoelectron, Ultraviolet Photoelectron, and Auger Electron Spectroscopy Spectra from First-Row Transition-Metal Nitrides: ScN, TiN, VN, and CrN, Surface Science Spectra. 7 (2000) 167–168. doi:10.1116/1.1360984.

[42] Bertóti, Characterization of nitride coatings by XPS, Surface and Coatings Technology. 151-152 (2002) 194–203. doi:10.1016/s0257-8972(01)01619-x.

[43] Glaser, S. Surnev, F.P. Netzer, N. Fateh, G.A. Fontalvo, C. Mitterer, Oxidation of vanadium nitride and titanium nitride coatings, Surface Science. 601 (2007) 1153–1159. doi:10.1016/j.susc.2006.12.010.

[44] Patscheider, N. Hellgren, R.T. Haasch, I. Petrov, J.E. Greene, Electronic structure of the SiNx/TiN interface: A model system for superhard nanocomposites, Physical Review B. 83 (2011). doi:10.1103/physrevb.83.125124.

[45] J.F. Moulder, W.F. Stickle, P.E. Sobol, K.D. Bomben, Handbook of X-ray Photoelectron Spectroscopy; Chastain, J, Perkin-Elmer Corp. , Eden Prairie, MN. (1992).

[46] N. Jiang, H.J. Zhang, S.N. Bao, Y.G. Shen, Z.F. Zhou, XPS study for reactively sputtered titanium nitride thin films deposited under different substrate bias, Physica B: Condensed Matter. 352 (2004) 118–126. doi:10.1016/j.physb.2004.07.001.

[47] Milosv, H.-H. Strehblow, B. Navinsek, M. Metikos-Hukovic, Electrochemical and thermal oxidation of TiN coatings studied by XPS, Surface and Interface Analysis. 23 (1995) 529–539. doi:10.1002/sia.740230713.

[48] N.C. Saha, H.G. Tompkins, Titanium nitride oxidation chemistry: An x-ray photoelectron spectroscopy study, Journal of Applied Physics. 72 (1992) 3072–3079. doi:10.1063/1.351465.

[49] F. Esaka, K. Furuya, H. Shimada, M. Imamura, N. Matsubayashi, H. Sato, A. Nishijima, A. Kawana, H. Ichimura, T. Kikuchi, Comparison of surface oxidation of titanium nitride and chromium nitride films studied by x-ray absorption and photoelectron spectroscopy, Journal of Vacuum Science & Technology A: Vacuum, Surfaces, and Films. 15 (1997) 2521–2528. doi:10.1116/1.580764.

[50] G.J. Leigh, J.N. Murrell, W. Bremser, W.G. Proctor, On the state of dinitrogen bound to rhenium, Journal of the Chemical Society D: Chemical Communications. (1970) 1661. doi:10.1039/c29700001661.

[51] G.J. Leigh, W. Bremser, X-Ray photoelectron spectroscopic studies of tertiary phosphine complexes of heavy transition metals, Journal of the Chemical Society, Dalton Transactions. (1972) 1216. doi:10.1039/dt9720001216.

[52] V.I. Nefedov, V.S. Lenenko, V.B. Shur, M.E. Vol'pin, J.E. Salyn, M.A. Porai-Koshits, Molecular nitrogen as a ligand. A study of dinitrogen complexes of transition metals, diazocompounds, and azides by x-ray photoelectron spectroscopy, Inorganica Chimica Acta. 7 (1973) 499–502. doi:10.1016/s0020-1693(00)94872-2.

[53] Chatt, C.M. Elson, N.E. Hooper, G. Jeffery Leigh, On the charge distribution in complexes, Journal of the Chemical Society, Dalton Transactions. (1975) 2392. doi:10.1039/dt9750002392.

[54] P. Brant, R.D. Feltham, X-ray photoelectron spectra of molybdenum dinitrogen complexes and their derivatives, Journal of the Less Common Metals. 54 (1977) 81–87. doi:10.1016/0022-5088(77)90128-x.

[55] N. Heide, J.W. Schultze, Corrosion stability of TiN prepared by ion implantation and PVD, Nuclear Instruments and Methods in Physics Research Section B: Beam Interactions with Materials and Atoms. 80-81 (1993) 467–471. doi:10.1016/0168-583x(93)96162-6.

[56] J.C. François, Y. Massiani, P. Gravier, J. Grimblot, L. Gengembre, Characterization and optical properties of thin films formed on TiN coatings during electrochemical treatments, Thin Solid Films. 223 (1993) 223–229. doi:10.1016/0040-6090(93)90525-t.

[57] Y.M. Shulga, V.N. Troitsky, Study of the surface of finely dispersed titanium nitride by x-ray photoelectron spectroscopy, Powder metallurgy (USSR) 10 (1979) 1-5.

[58] Y.M. Shulga, G.Y. Klyashchitsky, V.I. Rubtsov, E.N.Kurkin, V.N.Troitsky, Study of ultrafine powders from the Nb-N system by XPS and AES methods, Surface (USSR) 9 (1991) 67-77.

[59] V.I.Rubtsov, Y.M.Shulga, V.N. Troitski, E.N. Kurkin, A.A. Budanov Electron spectroscopy study of $Nb_{1-x}Ti_xN_{1-y}C_y$ ultrafine powders, Phys.Low-Dim. Struct. 12 (1995) 287-292.

[60] Y. Massiani, A. Medjahed, P. Gravier, L. Argème, L. Fedrizzi, Electrochemical study of titanium nitride films obtained by reactive sputtering, Thin Solid Films. 191 (1990) 305–316. doi:10.1016/0040-6090(90)90382-n.

[61] S.J. Tauster, Strong metal-support interactions, Acc. Chem. Res. 20 (1987) 389–394. doi:10.1021/ar00143a001.

[62] C.-J. Pan, M.-C. Tsai, W.-N. Su, J. Rick, N.G. Akalework, A.K. Agegnehu, S.-Y. Cheng, B.-J. Hwang, Tuning/exploiting Strong Metal-Support Interaction (SMSI) in Heterogeneous Catalysis, J. Taiwan Inst. Chem. Eng. 74 (2017) 154–186. doi:10.1016/j.jtice.2017.02.012.

[63] M.D. Ackermann, T.M. Pedersen, B.L.M. Hendriksen, O. Robach, S.C. Bobaru, I. Popa, C. Quiros, H. Kim, B. Hammer, S. Ferrer, J.W.M. Frenken, Structure and Reactivity of Surface Oxides on Pt(110) during Catalytic CO Oxidation, Physical Review Letters. 95 (2005). doi:10.1103/physrevlett.95.255505.

[64] Y. Feng, Q. Wan, H. Xiong, S. Zhou, X. Chen, X.I. Pereira Hernandez, Y. Wang, S. Lin, A.K. Datye, H. Guo, Correlating DFT Calculations with CO Oxidation Reactivity on Ga-Doped Pt/CeO2 Single-Atom Catalysts, J. Phys. Chem. C. 122 (2018) 22460–22468. doi:10.1021/acs.jpcc.8b05815.